# A Statistical Mixture-of-Experts Framework for EMG Artifact Removal in EEG: Empirical Insights and a Proof-of-Concept Application

Benjamin J. Choi[1], Griffin Milsap[2], Clara A. Scholl[2], Francesco Tenore[2], Mattson Ogg[2]

[1]*Harvard John A. Paulson School of Engineering and Applied Sciences, Cambridge, MA, United States of America*
[2]*Johns Hopkins University Applied Physics Laboratory, Laurel, MD, United States of America*

## Abstract

*Objective*

Effective control of neural interfaces is limited by poor signal quality. While neural network-based electroencephalography (EEG) denoising methods for electromyogenic (EMG) artifacts have improved in recent years, current state-of-the-art (SOTA) models perform suboptimally in settings with high noise. Since neural interfaces rely almost universally on some form of signal processing, algorithmic insights into denoising could enable improved EEG signal quality and support better machine control.

*Approach*

To address the shortcomings of current machine learning (ML)-based denoising algorithms, we present a signal filtration algorithm driven by a new mixture-of-experts (MoE) framework. Our algorithm leverages three new statistical insights into the EEG-EMG denoising problem: (1) EMG artifacts can be partitioned into quantifiable subtypes to aid downstream MoE classification, (2) local experts trained on narrower signal-to-noise ratio (SNR) ranges can achieve performance increases through specialization, and (3) correlation-based objective functions, in conjunction with rescaling algorithms, can enable faster convergence in a neural network-based denoising context.

*Main Results*

We empirically demonstrate these three insights into EMG artifact removal and use our findings to create a new downstream MoE denoising algorithm consisting of convolutional (CNN) and recurrent (RNN) neural networks. We tested all results on a major benchmark dataset (EEGdenoiseNet) collected from 67 subjects. We found that our MoE denoising model achieved competitive overall performance with SOTA ML denoising algorithms and superior lower bound performance in high noise settings.

*Significance*

These preliminary results highlight the promise of our MoE framework for enabling advances in EMG artifact removal for EEG processing, especially in high noise settings. Further research and development will be necessary to assess our MoE framework on a wider range of real-world test cases and explore its downstream potential to unlock more effective neural interfaces.

The supporting code to reproduce the figures and analyses in this manuscript can be found here:
https://drive.google.com/file/d/16onUqvfbDiUJs5-3Stpw1w1xEiHeCsfw/view?usp=sharing

## 1. Introduction

Poor signal quality remains a major barrier to high-performing non-invasive neural interfaces [1]. Quality issues stem from two main categories: system interference (e.g., sensor dislocation, breakage, etc.) and biological interference (e.g., contamination of EEG due to EMG artifacts). While the former category can be addressed via advances in hardware, solutions for the latter primarily rely on software-based algorithmic solutions [1-3] which are often framed as a blind source separation (BSS) problem.

Signal processing algorithms (canonical correlation analysis, or CCA; independent component analysis, or ICA, etc.) have been ubiquitous among neural interface applications over the last few decades, but these can fail to handle complex single-channel EEG-EMG interference [4-7]. Machine learning (ML)-based signal processing algorithms, however, possess the requisite expressivity to handle complex single-channel source separation tasks. While state-of-the-art (SOTA) signal filtration methods based on multi-layer perceptrons (MLPs), convolutional neural networks (CNNs), recurrent neural networks (RNNs), and transformers have begun outperforming traditional signal processing approaches, performance remains suboptimal in noisy environments [8-15]. Inherent difficulties in EEG denoising—especially in low signal-to-noise ratio (SNR) regimes with weak EEG signal power—continue to present challenges for deep learning (DL) models. Despite posting reconstruction correlation coefficients (CCs) above 0.9 in higher SNR regimes (i.e., >0 dB) [12-15], existing SOTA models have failed to demonstrate mean test CCs above 0.7 [12-15] at lower EEG-EMG SNRs (i.e., -7 dB) [12-15]. High noise settings require more exhaustive filtration to ensure a clean signal—leading to the undesirable filtration of target EEG signal components and reduced correlation between the filtered signal returned by the model and underlying ground truth [8-15].

Mixture of experts (MoE) approaches have achieved success in many ML tasks [16-17] including successful applications to EEG data [18-20]. MoE-based models use a routing mechanism to subdivide the input space and apply local expert models on each problem space region. MoE frameworks have been successfully applied to a number of classification tasks involving neural data [18-21], but these have yet to be formally applied to the EEG denoising problem; notably, applications of MoE for EEG artifact removal have been limited, owing to a lack of standardized benchmarks and appropriate training data [12]. The MoE approach is reliant on effective input space partitioning [16], and few input space-related or partitioning-relevant approaches have been developed for the EEG denoising problem [22].

In this study, we demonstrate three novel insights into EEG-EMG denoising, and use these insights to inform a proof-of-concept MoE model. Specifically, we show that:

1. EMG artifacts can be deterministically partitioned into quantifiable subtypes, and as a consequence, partitioning the denoising problem space by implicit EMG artifact type can improve filtration results in the downstream MoE model.

2. Subsetting the filtration problem space by inferred SNR can also lead to performance improvements by tailoring local MoE submodels to narrow SNR regions.
3. Correlation-based loss can enable faster convergence in neural network-based denoising models, and the corresponding loss of output scale can be remedied through specialized rescaling algorithms.

Preliminary formulations of both SNR-based partitioning and a specialized rescaling algorithm were included in our previous work [23] at a smaller scale, but are demonstrated more extensively herein. All three insights form the basis for our first-of-its-kind MoE model for EMG artifact filtration, which we develop and evaluate in this study. Given the track record of MoE at improving performance in diverse ML contexts [16-17, 24], we anticipated that—with the help of our partitioning insights—this MoE denoising model could improve artifact filtration performance relative to SOTA counterparts. We specifically hypothesized that the tailored training of local experts could help address the complexity of low SNR filtration cases—thereby addressing the shortcomings of existing SOTA models in high noise settings.

## 2. Methods

In order to evaluate the above approaches and assess the proposed MoE architecture, careful work was taken to determine an optimal deployment case. In the lineage of our prior work on EEG denoising [26], we surveyed numerous available EEG datasets [12, 25-27], taking into account the volume of available data, number and length of individual segments, availability of a potential interference source, and the potential for ground truth verification. After careful consideration, we selected the EEGdenoiseNet benchmark dataset for this study [12]. EEGdenoiseNet was chosen due to its comprehensive scale, incorporating data from five separate studies, optimization for ML artifact removal training, and inclusion of ground-truth data—qualities favorable to those of other open-source options. The dataset provides "4514 clean EEG segments" and "5598 muscular artifact segments" collected from 67 human participants at a 256 Hz digital sampling rate; the two-second segments enable "users to synthesize contaminated EEG segments with the ground-truth clean EEG" [12]. Furthermore, EEGdenoiseNet has been used extensively in recent years as a primary benchmark in the EEG denoising field [8-15]; the existence of published ML denoising algorithms trained on EEGdenoiseNet presents a robust baseline to evaluate our denoising model against the current SOTA [12-15]. An Nvidia A100 GPU (*Nvidia Corp., Santa Clara, CA, USA*) was used for training; end-to-end MoE training completed in <7 hours.

We used the EEGdenoiseNet dataset [12] to train and evaluate our proposed approaches. (The reader is referred to the original publication [12] for full details.) Contaminated signals were generated via a semi-synthetic EEG and EMG mixture method with hyperparameter selection calibrated to correspond to a conventional -7 dB to 2 dB SNR range for EMG-EEG denoising [12]. The employed semi-synthetic contamination method is commonplace in the neural BSS field and is well-validated as a proxy for real-world testing [8-15]. Our insights and MoE-based denoising efforts are situated in the context of this blind source separation formulation.

### 2.1 Partitioning by EMG Artifact Type

A key component of any MoE system is effective partitioning of the input space [16-17]. Our first partitioning-related insight relies on the notion that EMG artifacts can be heterogeneous in nature—and thus a portion of the problem space of contaminated EEG samples may be partitionable based on the type of EMG artifact present. Some sparse literature exists categorizing different types of EMG artifacts encountered in EEG data [28-29]; for example, Goncharova et al., 2003 identified distinct pattern classes of "noise-like" artifacts, "railroad cross-tie" artifacts, and "beta rhythm-like" artifacts—but categorical descriptions were largely qualitative. As the MoE setting demands deterministic, easily-quantifiable partitioning criteria in order to facilitate initial routing within the problem space, we propose a simple quantitative proxy metric for partitioning EMG artifacts in EEG data. Specifically, we used the variance of the EMG artifact itself at recording as a proxy for the "type" of the EMG artifact. While the original EMG dataset used in EEGdenoiseNet [30] made note of more granular categories of specific muscular origin, we propose this simple variance-based proxy to avoid overfitting to the idiosyncrasies of dataset-specific collection processes—and to hopefully enable more robust downstream generalization.

Using the relative EMG artifact variance at recording time as our partitioning criterion, we categorize our EMG artifacts into three distinct types of equal training set frequency: Type 1 (low variance at recording), Type 2 (mid variance at recording), and Type 3 (high variance at recording). Note that ground truth information for EMG variance at recording time does not exist in a real-world use case, as EMG artifacts are unlikely to be recorded in isolation from the EEG; variance at recording, however, is suitable for our training context—and the variance-encoded types need only be implicitly inferred at test time. We selected three classes of equal size as an initial attempt to balance the downstream MoE tradeoff between specificity and routing difficulty; further studies beyond our proof-of-concept demonstration could explore more granular EMG stratification strategies. As variance at recording is easily quantifiable, our EMG type-based partitioning satisfies our deterministic criterion. It is important to highlight, however, that (1) artifact variance at recording does not meaningfully translate to artifact variance in the semi-synthetic contamination problem, as the EMG artifact is normalized and scaled relative to the EEG signal to match the desired SNR level, and (2) the EMG artifact itself is only implicitly present in the contaminated signal (i.e., the input problem), as it is combined with EEG data. Thus, while EMG artifacts can be deterministically categorized by variance when recorded in isolation to help identify ground truth subtypes, "variance at recording" itself has no substantive meaning in the input space. Instead, it serves only as a measurable proxy for grouping fundamentally distinct EMG patterns.

In order to provide utility in an MoE context, our EMG artifact variance-based types must (1) correspond to meaningful distinctions requiring separate local experts, and (2) be reasonably distinguishable by a routing classifier. To assess these two criteria independent of any downstream MoE, we conduct the following two experiments. First, to assess the necessity of local experts and the implicit significance of the EMG subtypes themselves, we explore the "learnability" of EEG denoising with each EMG artifact type. If certain EMG types are easier to learn than others, then the EMG types likely constitute meaningful partitions in the EEG denoising problem space—and thus could offer utility in a downstream MoE context. Second, to assess whether the EMG type-based partitions are reasonably distinguishable, we train a CNN-based routing model to implicitly detect the different EMG artifact types within contaminated EEG data. The CNN-based routing model was subsequently employed to serve as a routing model for our downstream MoE system (see Section 2.4).

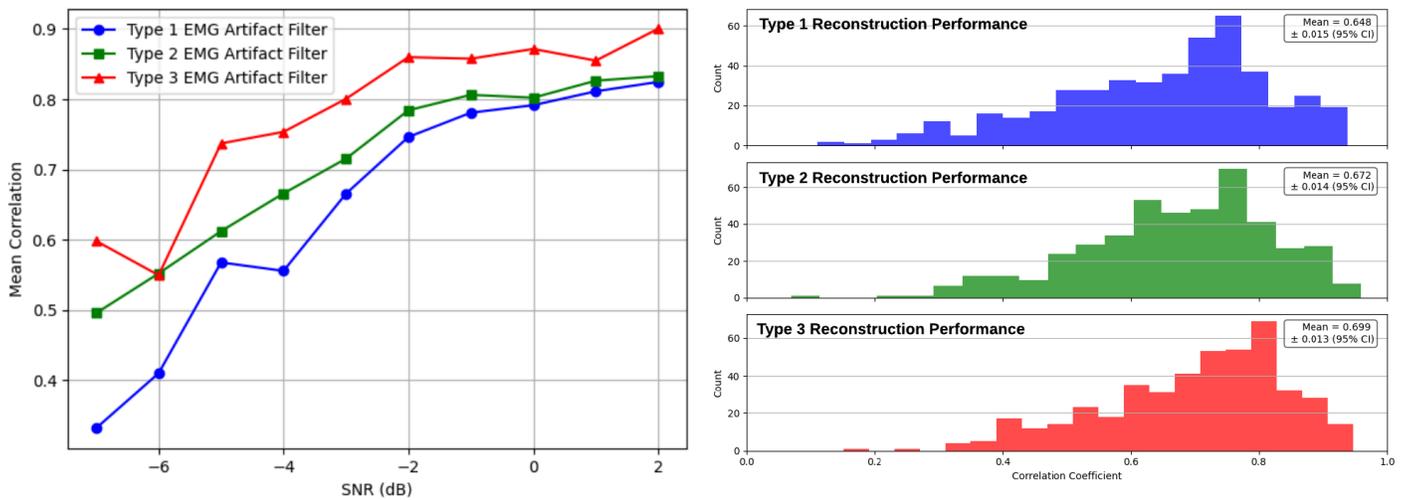

**Figure 1:** Validation performance of three identical CNN models trained and evaluated on separate problem space partitions based on EMG artifact type. We observe statistically significant differences in validation performance: contaminated signals of Type 1 (mean reconstructive correlation coefficient, or CC, with ground truth of 0.648±0.015) are hardest to filter, followed by Type 2 (CC of 0.672±0.014) and then Type 3 (CC of 0.699±0.013).

Using the EEGdenoiseNet semi-synthetic contamination protocol [12], we partitioned our training set EMG artifacts into three types and explored the relative difficulty of denoising using three identical models. Inspired by the past success of convolutional architectures in the denoising space [13, 23], we employ CNN-based denoising models throughout this study; our denoising models for this initial experiment on EMG type learnability consisted of three CNNs, each with ~2.1M parameters. After training each experimental model over 300 epochs with a batch size of 128 on a 9:1 train-validation split, we observed statistically significant differences in validation performance (at the 95% confidence level, with confidence intervals calculated as mean ± 1.96 × SEM) across the conventional -7 dB to 2 dB SNR range, as shown in Figure 1. Contaminated signals of Type 1 (i.e., containing EMG artifacts with the highest variance when initially recorded) were hardest to filter, with Type 2 next-most difficult and Type 3 easiest. Again, while variance itself has no visible meaning given that all artifact variances are rescaled at contamination-time, the observed empirical differences in learnability indicate that EMG variance at recording-time can serve as a meaningful proxy that can be used during training to partition different contaminated signals for routing.

The significance of our EMG types was further substantiated by our second experiment, in which we trained a 1.1M parameter CNN model to classify different semi-synthetically contaminated signals across the -7 dB to 2 dB SNR range based on EMG artifact type. Notably, this CNN classifier does not separate EEG and EMG before making a prediction—rather, it attempts to implicitly detect the type of EMG artifact contained within the contaminated signal. We found that our classifier achieved 72.35% test accuracy in distinguishing between Type 1, Type 2, and Type 3 input signals (with more granular classification results included in Figure 5). The classifier's ability to distinguish contaminated signals based on implicit artifact type serves as further empirical evidence that our variance-based proxy meaningfully partitions the EEG denoising problem space based on artifact type. Thus, this CNN-based classification model was used for routing to the downstream MoE models presented in Section 2.4 & Section 3.

## 2.2 Partitioning by SNR Level

In conjunction with partitioning our MoE system by EMG artifact type, we also investigated a complementary insight into the EEG denoising problem: whether partitioning contaminated signals by inferred SNR level can lead to elevated MoE performance by tailoring local experts to narrower SNR regions. This insight builds off our previous work [23], in which we presented a preliminary demonstration of SNR inference on discrete test cases and found that tailoring models to narrow SNR regions led to strong filtration results. In this study, we partition the continuous EEG SNR problem space into three regions: -7 dB to -4 dB (low SNR), -4 dB to -1 dB (mid SNR), and -1 dB to 2 dB (high SNR). We choose three regions in an attempt to balance granularity with routing difficulty; future work may explore more elaborate SNR-based partitioning schemes.

While assigning contaminated signals to local experts based on SNR can improve downstream performance, the process of routing input signals based on inferred membership in continuous SNR ranges poses challenges. To evaluate the viability of SNR-based routing—and in turn, the viability of a downstream MoE incorporating SNR partitioning—we trained a 1.1M-parameter CNN model to classify contaminated EEG segments drawn uniformly from a -7 dB to 2 dB range into one of the three aforementioned SNR tiers. Notably, this detection of SNR is implicit; the CNN classifier does not attempt to separate the EEG and EMG before making a prediction, but rather attempts to detect the SNR level directly from the contaminated input signal. We found that our SNR tier classifier was able to achieve 87.61% test accuracy in inferring the SNR level of contaminated signals, providing empirical evidence for the viability of SNR region-based partitioning. This CNN classifier was incorporated as a routing model into the downstream MoE system (see Section 2.4 & Section 3); more granular classification results are displayed in Figure 5 and Figure 6.

## 2.3 Correlation-Driven Training and Rescaling

In addition to developing novel approaches to partitioning EEG data based on EMG artifact subtypes and SNRs, our final insight is centered on neural network training objectives for the EEG denoising problem. This insight is loosely based on our previous work [23], in which we demonstrated a correlation-based objective function for training a denoising neural network. In our prior proof-of-concept demonstration [23], we presented a method to overcome the loss of appropriate scale associated with correlation-based loss functions (i.e., since models trained with correlation loss do not preserve the absolute scale of the original signal) by using a specialized scaling algorithm to recenter and rescale the denoised network output. Our previous rescaling algorithm achieved success by exploiting the correlation between the original and predicted signal, but required extensive built-in contingencies in order to avoid improperly rescaling various edge cases. In this study, we instead present a more generalizable approach to rescaling using a recurrent neural network (RNN). Specifically, in conjunction with the primary correlation-driven denoising model—in this case, a CNN—we train an RNN on an identical data partition using mean squared error (MSE) loss. At inference, we then recenter and rescale the correlated, out-of-scale CNN output to the same range as the corresponding RNN prediction. Empirically, we found that this RNN-based rescaling method enabled competitive performance in terms of temporal and spectral relative root mean squared error without any loss in correlation coefficient; more detailed specifications and results in a formal denoising context are discussed in Section 2.4 & Section 3.

The benefits of a correlation-based loss function as opposed to traditional MSE loss were initially noted in our previous study [23], in which we observed that correlation-driven training enabled high-fidelity filtration with a reduced compute burden during training [23]. Here, we expand upon our prior work by formalizing the benefits of correlation-driven training through both an initial experiment (Figure 2) and subsequent successful deployment in our MoE system. In our initial experiment, we found that an 8.5M-parameter denoising CNN trained with correlation-based loss achieved >10x faster convergence (using an NVIDIA A100 GPU for training) than an identical denoising CNN trained with MSE loss (Figure 2). Moreover, when combined with the aforementioned RNN rescaling algorithm using a supplemental 2.6M-parameter RNN, we found that our CNN and RNN-based denoising system achieved highly competitive filtration performance across both correlation and MSE-based metrics (see performance reports in Section 2.4 & Section 3) with a >10x reduction in training time.

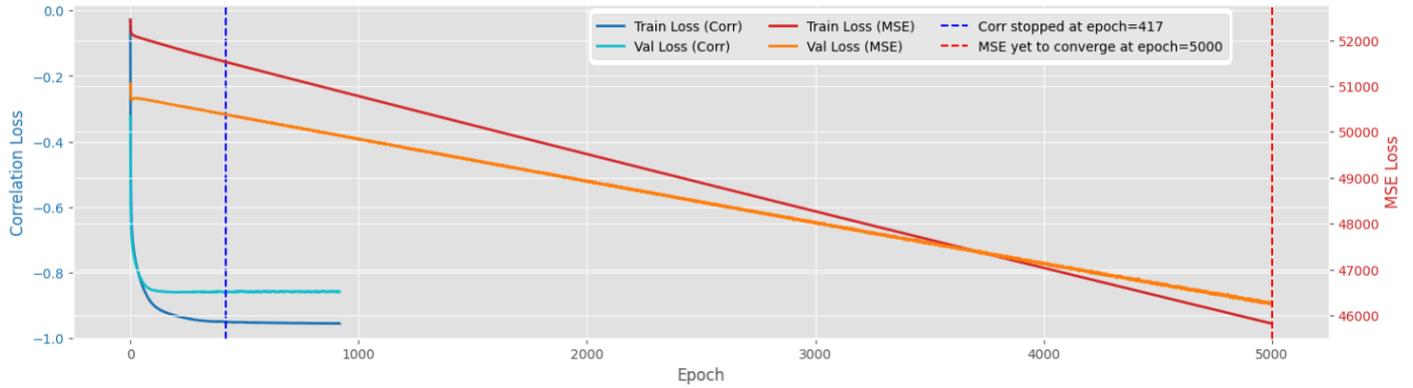

**Figure 2:** Denoising model convergence of two identical CNN architectures—one trained with correlation loss (*blue*) and the other trained with conventional mean squared error loss (*orange & red*). We found that employing correlation loss enabled a >10x reduction in training time while maintaining highly competitive filtration performance when combined with an RNN-based rescaling algorithm.

## 2.4 The Mixture-of-Experts Model

In line with our goal of demonstrating a successful MoE denoising model for filtering high noise EEG, we leveraged our three aforementioned insights—two with respect to partitioning (EMG type and SNR tier), and one with respect to correlation-driven rescaling (during training and for reconstructing model output)—to inform a novel MoE denoising system.

Our final MoE model consists of seven local experts, determined via a two-stage routing pipeline that partitions data along both EMG type and SNR tier. Specifically, two CNN routing classifiers (1.1M parameters each) operate in parallel: one routes input based on EMG artifact subtype, while the other classifies the SNR tier. Only one expert is invoked per input—there is no averaging across expert outputs. Each of the six experts covering low and mid SNR tiers consists of a 23.6M-parameter CNN trained with correlation loss in conjunction with a 2.6M-parameter RNN trained to rescale and reconstruct the denoised signal. The seventh expert—assigned to the high SNR tier regardless of EMG type—uses a standalone 4.7M-parameter RNN trained with MSE loss, motivated by previous observations [13] empirical findings that correlation-based CNNs underperform in low-noise regimes. All experts are trained independently (without need for serialization) on their respective partitions to best capture localized structure in each subset of the input space. The router is also trained separately and frozen during expert training. This modular training scheme simplifies optimization and allows us to tailor each expert to its target distribution. The full architecture of our final MoE model is depicted in Figure 3.

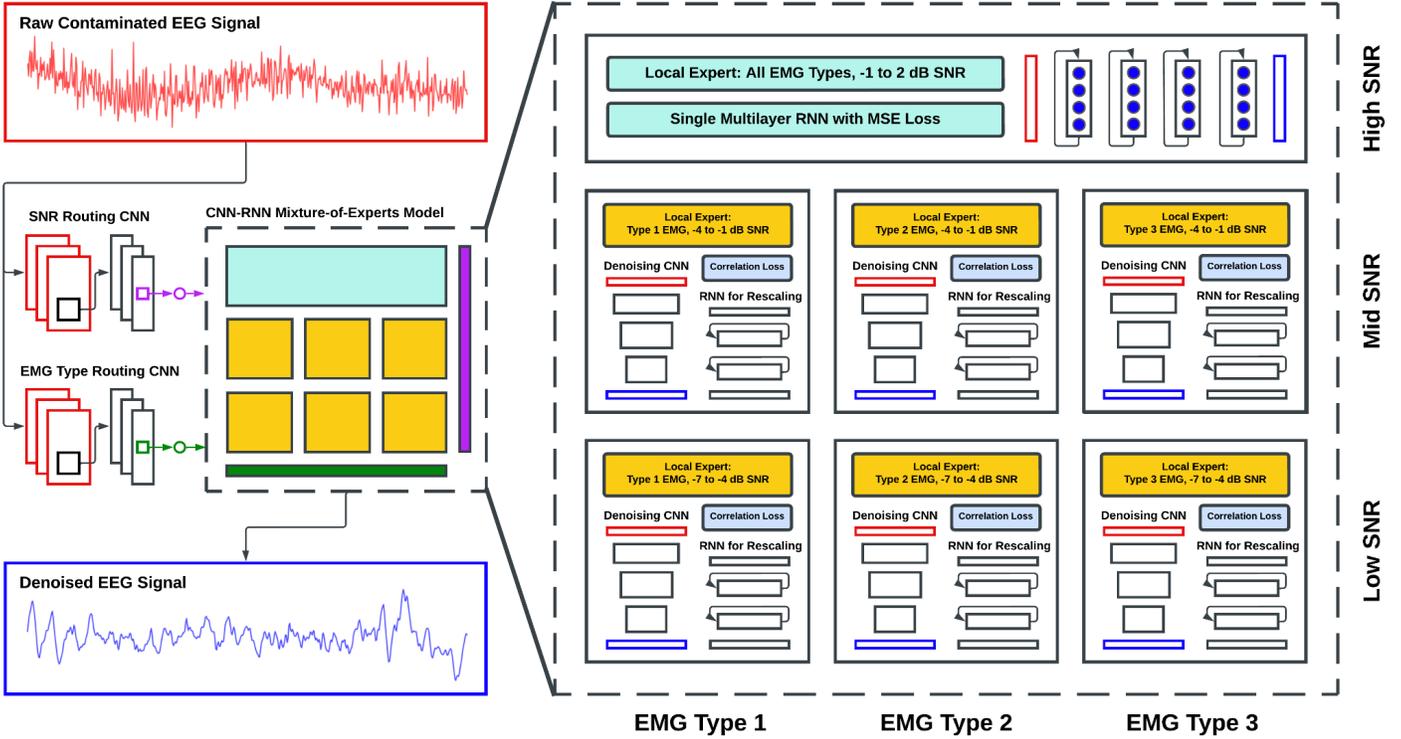

**Figure 3:** Diagram of final mixture-of-experts system. Contaminated signal (upper left) is fed into the two CNN routing classifiers, which predict corresponding EMG type and SNR tier. Based on the two routing predictions, one of seven local experts is invoked: either a CNN with correlation loss and supplemental RNN-driven rescaling for the six partitions at low and mid SNR, or a single RNN with MSE loss for the sole partition at high SNR. The selected local expert outputs the final denoised EEG signal (lower left).

The specific seven-expert MoE structure and partitioning scheme was determined via methodical ablation analysis. Specifically, we explored three potential MoE structures: MoE using only EMG type partitioning, MoE using only SNR partitioning, and MoE using both EMG type partitioning and SNR partitioning. Taking inspiration from past success with CNN denoising architectures [13] and loosely building off our past work on convolutional EEG filtration models [23], we employed 23.6M-parameter CNN local experts for all MoE experimentation in conjunction with local 2.6M-parameter rescaling RNNs. The implications of model footprint are discussed in Section 4.

**Table 1:** Results of ablation experiment for determining optimal mixture-of-experts structure.

| Model | High Noise (C-T-S) | Overall Mean (C-T-S) |
|---|---|---|
| **EEGdenoiseNet Baseline (RNN) [12]** | 0.55-1.03-1.00 | 0.81-0.57-0.53 |
| **MoE: EMG Type Partition** | 0.64-0.74-0.74 | 0.80-0.54-0.56 |
| **MoE: SNR Partition** | 0.72-0.67-0.70 | 0.85-0.49-0.51 |
| **MoE: EMG and SNR Partition (9x)** | 0.69-0.72-0.65 | 0.81-0.53-0.55 |
| **MoE: EMG and SNR Partition (7x)** | 0.75-0.68-0.63 | 0.83-0.51-0.54 |

*C-T-S denotes correlation coefficient, temporal RRMSE, and spectral RRMSE, respectively.*
*High Noise denotes mean performance at lower SNR bound (i.e., -7 dB).*

Across our examined MoE variants, we found that our MoE structures performed well in high noise settings (i.e., -7 dB) and remained competitive with published baselines across the entire -7 dB to 2 dB range (see Table 1). However, we found that combining three-tier SNR routing with three-tier EMG type routing to create a nine-partition MoE did not lead to increased performance; empirically, we observed that partitioning by EMG type only improved filtration accuracy in higher noise settings (i.e., low or mid SNR tier). In light of this observation, we evaluated a fourth and final MoE model that partitions the problem input space into seven parts: three EMG partitions at low SNR, three EMG partitions at mid SNR, and a single partition at high SNR. We found that this fourth MoE achieved the strongest performance in high noise settings (see Table 1); this seven-partition MoE thus became the final MoE system presented in this study.

### 3. Results

Our final seven-partition MoE model was evaluated per the semi-synthetic contamination protocol given in the original dataset description [12] across the entire EEGdenoiseNet dataset using a 9:1 train-test split. The model was trained and tested on a continuous -7 dB to 2 dB SNR range; in addition to evaluating overall performance, we provide additional analyses of performance within the lower SNR range (i.e., -7 dB) in order to validate our original goal of a successful MoE system for high noise EEG filtration. We selected three performance metrics—correlation coefficient with ground truth (CC), temporal relative root mean squared error (TRRMSE), and spectral relative root mean squared error (SRRMSE)—in order to strictly align with the evaluation metrics used in prior studies on EEG denoising [12-15] with EEGdenoiseNet.

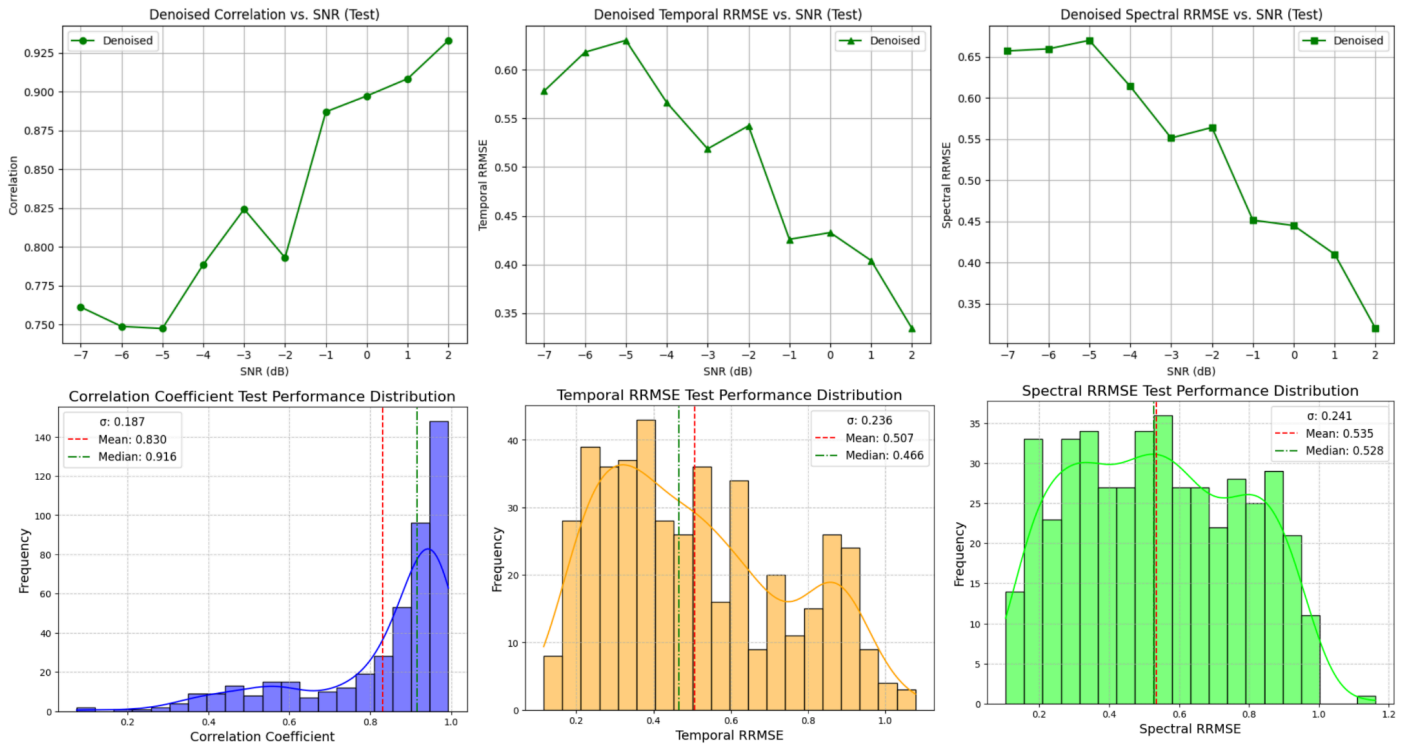

**Figure 4:** Test performance of the final mixture-of-experts system across the -7 dB to 2 dB range; binned scatter plots are per field convention, while distribution plots are added below to convey distribution information and statistical power. Three metrics are evaluated (from left to right): correlation coefficient with ground truth (mean: 0.830, median: 0.916, standard deviation: 0.187), temporal relative root mean squared error (mean: 0.507, median: 0.466, standard deviation: 0.236), and spectral relative root mean squared error (mean: 0.535, median: 0.528, standard deviation: 0.241). The mean test sample scatter plots (upper panel) depict

performance binned to the nearest integer SNR level; the lower panels depict histograms of individual test sample performance. A comparison with published SOTA models is included in Table 2; specific results are further discussed in Section 4.

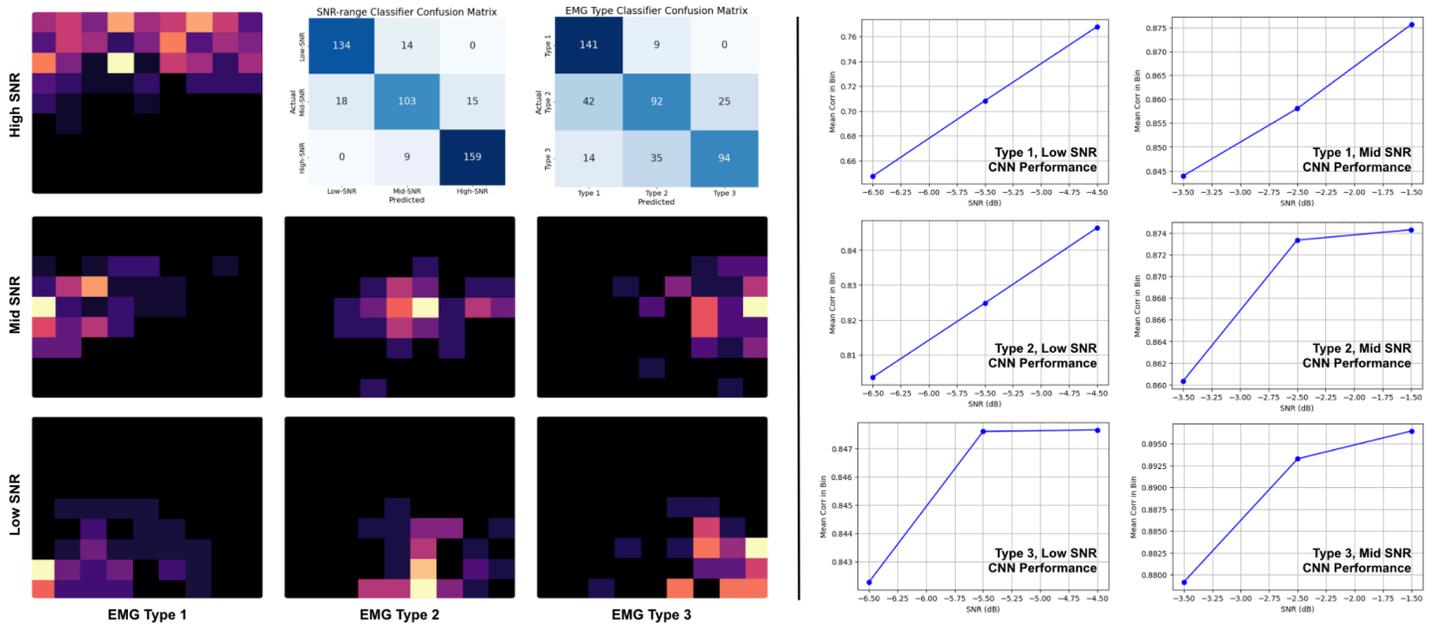

**Figure 5:** Further insights into the final mixture-of-experts system, with EMG type and SNR tier routing performance displayed on the left, and validation correlation loss performances displayed on the right. The seven heatmap plots (left) depict the ground truth distribution of test samples ultimately routed to each local expert (with EMG type as measured by initial variance at recording and SNR tier both converted to a nine-bin spectrum); brighter colors denote more frequent routing from individual bins. Two confusion matrices for the SNR range classifier and the EMG type classifier are also included on the left panel. The right panel depicts binned validation correlation loss for each of the six correlation-driven CNNs; local expert validation performance is ~10% higher than test performance as the validation sets assume perfect routing.

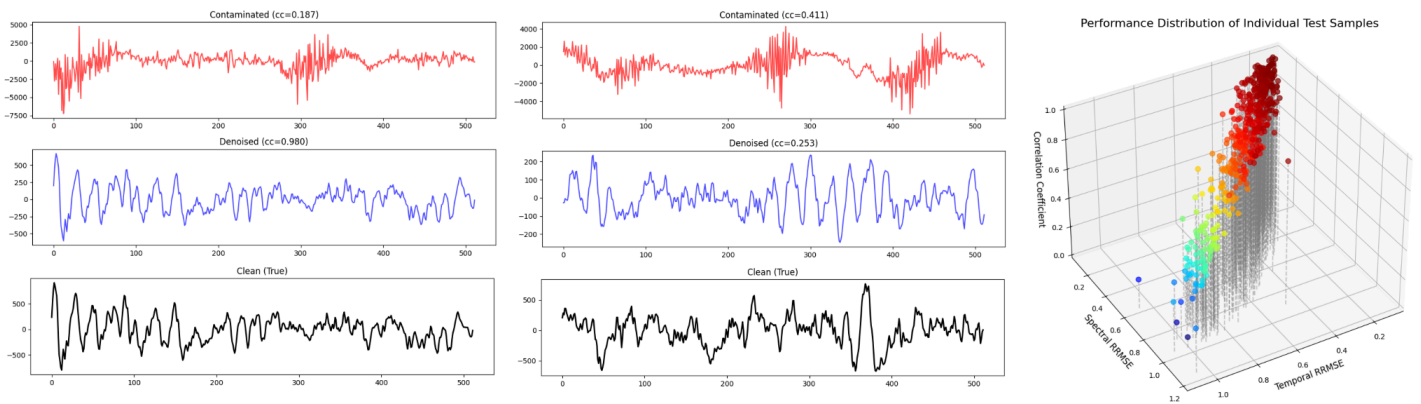

**Figure 6:** Further depictions of final MoE system performance—the best recorded test case (from top to bottom: contaminated, denoised, ground truth) is depicted on the left panel, while the worst recorded test case is depicted on the central panel. A 3D histogram of individual test sample performance (with axes of TRRMSE, SRRMSE, and CC) is depicted on the right panel.

We recorded a mean test latency of 0.215 seconds (with a standard deviation of 0.075s) for the end-to-end MoE system on an AMD EPYC CPU (*Advanced Micro Devices, Inc., Santa Clara, CA, USA*). Table 2 includes a comparison between the results reported by published SOTA models on EEG filtration with EEGdenoiseNet (all across the standard -7 dB to 2 dB range) and our final MoE system; we compare existing studies across the three aforementioned metrics at both high noise (i.e., -7 dB) and across the overall -7 dB to 2 dB range. All results in Table 2 are reported or best estimated from provided documentation; further analyses are included in the ensuing Discussion.

Table 2: Comparison with state-of-the-art denoising models across three performance metrics.

| Model & Year | High Noise (C-T-S) | Overall Mean (C-T-S) |
|---|---|---|
| RNN (2021) [12-13] | 0.55-1.03-1.00 | 0.81-0.57-0.53 |
| Novel CNN (2021) [13] | 0.69-0.72-0.65 | 0.86-0.45-0.44 |
| EEGDnet (2022) [14] | 0.53-0.92-0.85 | 0.73-0.68-0.63 |
| EEGDiR (2024) [15] | 0.46-1.00-0.90 | 0.81-0.53-0.50 |
| Mixture of Experts (2025) [Ours] | 0.75-0.68-0.63 | 0.83-0.51-0.54 |

*C-T-S denotes correlation coefficient, temporal RRMSE, and spectral RRMSE, respectively.*
*High Noise denotes mean performance at lower SNR bound (i.e., -7 dB).*

In short, our MoE denoising system achieved reasonably competitive overall performance (Table 2) across the entire -7 dB to 2 dB range. Our MoE model outperforms EEGDnet on all three metrics, outperforms both the top benchmark RNN [12-13] and EEGDiR [15] in terms of CC and temporal RRMSE, but falls short of Novel CNN [21] on all three metrics. However, our MoE model addresses the shortcomings of other systems in high-noise settings—our MoE outperforms all four models across all three metrics at poor SNR levels (-7 dB).

## 4. Discussion

Overall, as seen in Table 2, our MoE system's strong performance in high-noise EEG regimes offers a notable comparative advantage; the data provide preliminary validation for our original hypothesis that the more granular partitioning at low SNR—coupled with EMG subtype partitioning—can improve performance in high noise environments through more tailored local experts. We observe, however, some performance tradeoffs in other scenarios, as reflected in the less dominant overall denoising performance (Table 2, Figure 4). In comparison with existing SOTA models [12-15], the binned correlation performance plot in Figure 4 is not monotonically increasing, nor are the binned temporal and spectral RRMSE plots in Figure 4 monotonically decreasing. These apparent fluctuations in test performance stem from the detrimental consequences of routing misclassification associated with SNR and EMG type-based partitioning. Given perfect routing, local expert performance (see Figure 5) is indeed monotonically increasing with SNR and is ~10% higher across performance metrics than the test performance in Figure 4; the consequences of misrouted test samples particularly manifest in the relatively weaker test performance in the -6 to -2 dB range. Still, the overall worst case binned performance (0.75 CC) of our MoE across the -7 dB to -2 dB test range may provide its own relative advantages compared to existing SOTA models for use cases that require mean denoised EEG to exceed certain quality thresholds—or more general use cases with high levels of anticipated noise. Notably, our final MoE system also achieved strong median test performance (0.916 CC, 0.466 TRRMSE, 0.528 SRRMSE), with all three median performance metrics exceeding mean performance. While median performance data is less commonly reported in SOTA literature, none of the SOTA benchmarks reported in 2024 by EEGDiR [15] exceeded overall median test performance of 0.9 CC, indicating that our MoE may compare relatively favorably in terms of median performance. This strong median test performance may again be explained by the nature of input space partitioning, in which downward pressure is exerted on mean performance relative to median due to a minority of misrouted samples.

We ran our mean latency test over 1,000 trials on a CPU in order to simulate a relatively lightweight computing environment. While the aforementioned mean CPU latency of 0.215 seconds may be prohibitive for some demanding real-time applications, the independent routing models and discrete local expert structure may offer avenues for parallelization; inference could be further accelerated via GPU-specific optimizations. MoE model footprint is relatively large, consisting of two 1.1M-parameter routing models, six 23.6M-parameter CNNs, six 2.6M-parameter rescaling RNNs, and one 4.7M-parameter RNN for high SNR. As a result, in contrast with our previous work [23], our proof-of-concept MoE system may not be as tractable in resource-constrained settings. The EEG samples themselves, however, are only routed through one of the seven local experts, and so the effective parameter size at inference (i.e., the maximum combined model parameter count that an individual sample can be routed through) remains around 28.3M parameters; for context, the Novel CNN model [13] contains an estimated 58.7M parameters as best extrapolated by provided code. However, model footprint remains a notable drawback of our MoE approach.

While our study makes use of the conventional semi-synthetic validation techniques employed by prior work in the EEG denoising field [8-15] and includes—through EEGdenoiseNet—data from 67 total human participants, it is important to note that our MoE study constitutes a preliminary demonstration of viability. While beyond the scope of this study, future validation of our MoE system will likely require deployment in real-world neural interfacing use cases. Moreover, future work may explore more targeted forms of EMG type categorization beyond the contextual semi-synthetic proxy method used herein, as well as more elaborate insight-driven partitioning schemes beyond our seven-partition MoE structure.

---